\documentstyle[prl,aps,multicol,epsf,rotate]{revtex}
\draft
\tighten
\title
{\Large\bf Force-Extension Relation and Plateau Modulus for Wormlike Chains}
\author{Klaus Kroy and Erwin Frey}
\address{
Institut f\"ur Theoretische Physik, Technische Universit\"at M\"unchen, \\
James-Franck-Stra\ss e, 85747 Garching, Germany}

\begin{document}

\date{to appear in PRL}
\maketitle

\begin{abstract}
  We derive the linear force-extension relation for a wormlike chain of
  arbitrary stiffness including entropy elasticity, bending and thermodynamic
  buckling. From this we infer the plateau modulus $G^0$ of an isotropic
  entangled solution of wormlike chains. The entanglement length $L_e$ is
  expressed in terms of the characteristic network parameters for three
  different scaling regimes in the entangled phase. The entanglement transition
  and the concentration dependence of $G^0$ are analyzed. Finally we compare
  our findings with experimental data.
\end{abstract}

\begin{multicols}{2}
  
  
  Recently there has been increasing interest in biological material
  research~\cite{Geilo95}. The physical properties of colloids, liquid crystals
  and macromolecular networks are of prime importance for the structure and
  function of biological entities such as cells and muscles.  On the other hand
  biology provides physicists with some of the most pertinent model systems to
  test their theories of soft matter \cite{nel95}. Among these systems we will
  concentrate on macromolecular networks here. These networks may be assemblies
  of relatively flexible (DNA), semiflexible (actin) or rigid (microtubuli)
  molecules or complex compound systems as in the case of the cytoskeleton of
  erythrocyte plasma membranes or in the glass body of the eye \cite{sac94}.
  Especially the static and dynamic rheological properties of actin networks
  are crucial for an understanding of the mechanical stabilization and the
  motility of cells.
  
  Solutions of flexible polymers at high concentration are known to exhibit
  spectacular features in both their mechanical response and molecular
  relaxation \cite{fer80,doi92}, which are commonly attributed to the
  topological constraints due to the uncrossability of the polymers
  (entanglement).  For time scales shorter than a characteristic time $\tau$
  the response of a solution of high molecular weight polymers to a periodic
  perturbation is elastic over an extended frequency interval (``rubber
  plateau'') and resembles that of a permanently crosslinked network or gel. It
  is commonly agreed that polymeric liquids form a temporary network, where
  {\em entanglements} play a similar role as permanent crosslinks in gels.
  There have been several attempts to derive this successful phenomenological
  concept from other characteristic parameters of a polymer solution and thus
  gain some understanding of the underlying microscopic mechanisms
  \cite{col90,schweizer}. Over the last few years experiments with actin
  \cite{mul91,jan94} revealed that the above qualitative picture also holds for
  semiflexible polymers. However, the theoretical understanding of entangled
  solutions of semiflexible polymers is much less developed than in the
  flexible case, and experimental data are often interpreted within the
  theoretical framework established for flexible coils or rigid rods
  respectively \cite{jan94,zan95}.

  In this letter we develop some basic concepts for a general theory of
  rheology of isotropic entangled solutions of {\em wormlike chains}. The
  wormlike chain \cite{sai67} is the minimal model of an ideal (i.e.  non
  self-avoiding) polymer of arbitrary stiffness.  In contrast to the fractal
  Gaussian chain model \cite{doi92}, which can serve only as an effective large
  scale model for rather flexible polymers, the wormlike chain model also
  faithfully reproduces the intrinsic stiffness of real polymers. The polymer
  is represented as a differentiable space curve $\mbox{\boldmath $R$}_s$ with
  its statistical properties determined by the effective free energy
\begin{equation}
  \label{worham}
  H \left( \{ \mbox{\boldmath $R$}_s \} \right) = \frac \kappa2 \int_0^L\!\!ds
  \left( \frac{\partial^2\mbox{\boldmath $R$}_s}{\partial s^2}\right)^{\!\!2}
    \, .
\end{equation}
A central feature of this model is the inextensibility of the chain leading to
the rigid constraint $|\partial \mbox{\boldmath $R$}_s/\partial s|=1$, which
additionally has to be imposed on the contour.  Due to the mathematical
complications resulting from this constraint only few of the statistical
properties of the wormlike chain can be extracted analytically, the most
prominent being the mean square end-to-end distance $\langle \mbox{\boldmath
  $R$}^2_L \rangle = L^2f_D(L/\!L_p)$, with the Debye function
$f_D(x):=2(x-1+e^{-x})/x^2$.  (The persistence length $L_p$ is related to the
bending modulus $\kappa$ by $\kappa=L_pk_BT$.) For large $L/\!L_p$ this reduces
to the power law $\langle \mbox{\boldmath $R$}_L^2\rangle =2L_pL$
characteristic of a random walk of step length $2L_p$.  Note that the model
does not reproduce the swelling of real flexible coils, since the
self-avoidance of the chain is neglected in the above effective free energy,
Eq.\ (\ref{worham}). Therefore, the model is restricted to solutions with mesh
size $\xi_m$ not much larger than the persistence length $L_p$ of the polymers,
in which case self-avoidance effects can safely be neglected.

\begin{figure}
  \narrowtext \hspace{0.005\columnwidth} \epsfxsize=0.9\columnwidth
  \epsfbox{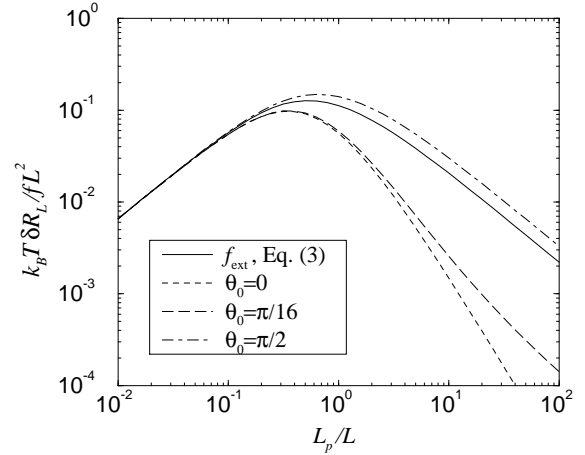}
  \caption{The deformation $\delta\!R_L$ of a wormlike chain of given length 
    $L$ to leading order in the applied force $f$ as a function of the
    persistence length $L_p$. See explanation in the text.}
   \label{cumplot}
\end{figure}

As a first step towards an understanding of the macroscopic viscoelastic
properties of an entangled network of wormlike chains one has to understand the
elastic properties of a single wormlike chain. The linear force-extension
relation of a wormlike chain is obtained by the following argument. Consider a
wormlike chain with one end clamped at fixed orientation at the origin. Apply a
weak force $f${\boldmath $n$} (directed along the unit vector {\boldmath $n$})
to the other end \cite{comple}. The configurational distribution
function is then modified by a Boltzmann factor $\exp(f\mbox{\boldmath
  $nR$}_L/k_BT)$.  The extension $\delta\! R_L:=\mbox{\boldmath $n$}(\langle
\mbox{\boldmath $R$}_L\rangle_f-\langle\mbox{\boldmath $R$}_L\rangle)$ in the
direction of the applied force to first order $f$ is given by the linear
extension coefficient $\tilde f^{-1}_{\theta_0}:=\partial\delta\!R_L/\partial
f|_{f=0}$,
\begin{equation} 
 \label{ext}
 \tilde f^{-1}_{\theta_0} = \int\!\! ds\!\! \int\!\! ds' \, \langle
 \cos\theta_s\cos\theta_{s'} \rangle - 
\left(\int\!\!ds\, \langle \cos\theta_s \rangle \right)^2 .
\end{equation}
By $\theta_s$ we denote the tilt angles of the tangents of the polymer contour
with respect to {\boldmath $n$}. The thermal average is to be taken under the
constraint that the angle $\theta_0$ at the clamped end is kept fixed. Standard
methods \cite{sai67} yield for $\tilde f^{-1}_{\theta_0}$ the dashed curves in
Fig.~\ref{cumplot}. In general, Eq.~(\ref{ext}) predicts a polymer of contour
length $L$ to appear more floppy if $L_p\simeq L$ than in the high temperature
limit ($L_p\to0$) and the low temperature limit ($L_p\to\infty$), when it
contracts to a little ball or becomes a rigid rod respectively. In the flexible
limit, where the chain becomes an isotropic random coil, all curves fall
together and reproduce entropy elasticity.  But for stiff chains, as a
consequence of the chain anisotropy, the force-extension relation depends
strongly on the value of $\theta_0$.  Obviously, $\theta_0=0$ is an exceptional
case. Whereas for all other angles $\theta_0$ the ultimate asymptotic form of
$\tilde f_{\theta_0}$ in the stiff limit is $\kappa/\!L^3$, at $\theta_0=0$ the
force coefficient becomes $\tilde f_0 \simeq \kappa^2\!/k_BTL^4$; i.e.\ it is
second order in the bending modulus and diverges at low temperatures $T$. The
latter result was previously obtained in Ref.\ \cite{mac95}. Note that
$\theta_0$ is the angle between the applied force and the average orientation
of $\mbox{\boldmath $R$}_L$, i.e.\ for $\theta_0=0$ the force is parallel to
$\mbox{\boldmath $R$}_L$ on average.  Especially for $T=0$ the force is pulling
or pressing on a rigid rod along its axis. In this limit the above expansion of
the Boltzmann factor breaks down and we encounter the so called Euler buckling
instability, i.e., the force-extension relation becomes highly nonlinear and
the force coefficient in linear response does not exist. This situation is well
known for foams and other cellular materials \cite{gib88}. If we require
$fR_L\ll k_BT$, the buckling instability is evaded by thermal undulations, and
we find a linear contribution to the force-extension relation (`thermodynamic
buckling').  But with decreasing temperature the volume fraction
$k_BTL/\!\kappa$ (stored thermal energy over bending energy) occupied by the
thermal undulations vanishes, and ultimately there remain no more undulations
to be bent or pulled out, hence the divergence of $\tilde f_0$ with $T^{-1}$.
We take as the force coefficient $\tilde f$ of a `general strand' of length $L$
in a random network the average of $\tilde f^{-1}_{\theta_0}$ over all
orientations $\theta_0$ \cite{kre95}:
\begin{equation}
  \label{deltaR}
  \tilde f^{-1}=L^2f_{\rm ext}(L/\!L_p)/k_BT\, ,
\end{equation}
with $f_{\rm ext}(x):=(2x-3+4e^{-x}-e^{-2x})/3x^2$ (the extension function).
This result is also shown in Fig.~\ref{cumplot}. In the stiff limit ($L\ll
L_p$) it reduces to $\tilde f \simeq \kappa/\!L^3$.

Now, to relate the force-extension relation of the general strand to the
observed elastic modulus of a polymer network in the rubber plateau regime we
proceed in close analogy to solid state physics.  In the harmonic approximation
the elasticity tensor {\bf E} of a monatomic Bravais lattice is written as
$E_{ijkl}=-\sum_{\{R\}}R_iD_{jl}(\mbox{\boldmath $R$})R_k/2V$, with $V$ being
the volume of the primitive cell, $\mbox{\boldmath $R$}$ the lattice vectors
and {\bf D} the matrix of second derivatives of the interaction potential with
respect to lattice displacements.  For an isotropic entangled polymer solution,
we take the analogue of the primitive cell to be an entanglement volume $V_e$
and the analogue of the primitive vectors to be the average distance $\xi_e$
between adjacent entanglements (in the embedding space).  This scaling argument
suggests that the storage modulus in the plateau regime should (up to a
constant factor depending on the strain geometry) be given by $G^0\simeq
c_e\tilde f_e \xi_e^2$. Here $\tilde f_e$ is the force coefficient of polymer
sections of length $L_e$ between adjacent entanglements and $c_e\simeq
V_e^{-1}$ their concentration. (As far as self-avoidance effects can be
neglected, $L_e$ and $\xi_e$ are related by the Debye function.) The situation
may be visualized by external forces acting on contour elements distributed
with an average spacing $L_e$ along the polymer.  Inserting $\tilde f$ from
Eq.\ (\ref{deltaR}) with $L_e$ substituted for $L$ into the formula for $G^0$
we finally arrive at the following explicit expression for the plateau modulus
of an entangled solution of wormlike chains,
\begin{equation}
  \label{G0}
  G^0 \simeq  c_ek_BT 
  \frac{f_D(L_e/\!L_p)}{f_{\rm ext}(L_e/\!L_p)}\simeq 
    \left\{
    \begin{array}{ll} 
      c_ek_BT & (L_e\gg L_p) \\ 
      c_e\frac\kappa{L_e} & (L_e\ll L_p)\, .
    \end{array}
    \right. 
\end{equation} 

The entanglement length $L_e$ is obviously the crucial quantity in Eq.\ 
(\ref{G0}). In the literature several scaling ideas \cite{col90} have been
reported on how $L_e$ and $\xi_e$ may be derived from the known static
properties of a flexible polymer network. Note however that Eq.\ (\ref{G0})
holds independently of such considerations. For a homogeneously crosslinked
gel of semiflexible or rod-like polymers $\xi_e$ can essentially be identified
with the mesh size $\xi_m$ of the network. In this case Eq.~(\ref{G0}) predicts
$G^0\propto \kappa c^2$ in the stiff limit. We conjecture that for a solution
of wormlike chains of arbitrary stiffness one has to distinguish three
different regimes. We will treat the limiting cases of scale invariant chain
structure -- i.e.\ a virtually Gaussian or straight conformation respectively
-- in a very similar manner. The breaking of scale invariance due to Eq.\ 
(\ref{worham}) gives rise to an intermediate regime for chains with $L\approx
L_p$, which will be discussed subsequently.

For a weakly bending contour we have from Eq.\ (\ref{worham}) the scaling
relation $R_L^{\perp2}=2L^3/3L_p$ for the transverse amplitudes $R_L^\perp$ of
the largest bending undulations. If these amplitudes are smaller than the mesh
size $\xi_m$ of the surrounding polymer network, i.e.\ $\xi_m^2>2L^3/3L_p$,
then the bending undulations are not substantially perturbed and are supposed
to be rather irrelevant to the question of entanglement. In this case we should
thus be allowed to represent the polymers as straight (but not rigid) ``rods''
in our derivation of $L_e$. In the opposite extreme ($L,\xi_m\gg L_p$) of a
strongly coiled polymer conformation the polymer may be represented by a
fractal curve (or a freely jointed chain of ``blobs'', if screened
self-avoidance is to be included). We feel that the flexible case has been
described successfully earlier \cite{kav87} and will adapt this approach to
straight rods now. It is based on the crucial observation that polymer ends are
not contributing efficiently to long-lived entanglements.  Namely, if
entanglements would depend on dangling ends, they could not be long-lived as
compared to unperturbed, free fluctuations of the polymer and hence could not
give rise to a rubber plateau.  To be specific, we assume that an entanglement
requires a sufficient number of non-end neighboring polymer segments that on
average restrict the lateral degrees of freedom of a test chain.  Consider a
sphere of radius $\xi_e$ around such a mean entanglement point. Then for a
given monomer concentration $c$ and volume fraction $cv$ the excluded volume in
this ``primitive cell of entanglement'' of volume $V_e = 4 \pi (\xi_e/2)^3 / 3$
is given by $c v V_e$.  In order to achieve an entanglement one requires that a
certain amount of polymer material, ${\cal C}\pi (a/2)^2 L_e$, is contained in
the test volume.  Here $a$ denotes the lateral diameter of the polymer. The
quantity ${\cal C}$ is a geometry factor which measures the amount of polymer
material in the test volume, if $K$ polymers cross the sphere around the test
chain \cite{footn}. We determine $L_e$ by equating the excluded volume (reduced
by the contribution coming from free ends) with the volume of polymer material
needed for an entanglement
\begin{equation}
  \label{ansatz}
  cv\left( 1-\frac{L_e}L\right)=\frac{3a^2L_e{\cal C}}{2\xi_e^3}.
\end{equation}
The implicit equation Eq.\ (\ref{ansatz}) can be solved analytically in the
random coil limit \cite{kav87} and for a straight conformation. For the latter
we find
\begin{equation}
  \label{Le}
  L_e=\frac L3\left\{ 1-2\sin\left[\frac13 \arcsin\left(1-\frac{27 L_e^{\infty
            2}}{2L^2}\right)\right]\right\} \, ,
\end{equation}
where $L_e^\infty := a\sqrt{3{\cal C}/2cv}\simeq\sqrt{\cal C} \xi_m$ is the
entanglement length in the limit of infinitely long molecules. Eq.\ (\ref{Le})
describes an entanglement transition of the polymer solution characterized by a
cusp singularity of the entanglement length $L_e$ as a function of the polymer
length $L$ or the volume fraction $c v$. The phase boundary between the
entangled and the disentangled regime is given by either of the two equations
$c^\star(L) = 27 {\cal C} {\bar c} (L) / 4$, $L^\star (c) = 3 \sqrt{3}
L_e^{\infty} (c) / 2$, where $\bar c$ denotes the geometrical overlap
concentration ${\bar c} v = 3 a^2/ 2 L^2$.  The value of the entanglement
length and the contour length at the cusp singularity are related by
$L_e^\star= 2 L^\star / 3$. The above results have some important
consequences on the rheological properties of semiflexible polymer solutions.
Upon taking the experimental value for the geometric constant ${\cal
  C} = 9.1$ \cite{kav87} of flexible polymers to be a universal quantity (also
valid in the rod-limit) one estimates that the critical polymer length
for the solution to show entanglement has to be about eight times the mesh
size. Accordingly, the critical concentration $c^\star$ is predicted to be
almost two orders of magnitude larger than the overlap concentration ${\bar
  c}$, $c^\star / {\bar c} = 27 {\cal C} / 4$. In the intermediate
concentration regime, $c^\star > c > {\bar c}$, there is already a significant
overlap of the semiflexible polymers but no long-lived entanglements leading to
a rubber plateau regime.  In this disentangled phase the magnitude of the
storage modulus is supposed to show a linear concentration dependence.

Finally we comment shortly on the intermediate case of a network of wormlike
chains with a mesh size $\xi_m$ smaller than the persistence length $L_p$ and
the amplitudes of the largest bending undulations $R_L^\perp$. To distinguish
it from the case discussed above, which could be called the ``rod-like''
regime, we will address it as the ``snake-like'' regime. It is characterized by
the property that all bending undulations with wavelength longer than a
critical wavelength, the ``deflection length'' \cite{odi83} $\lambda\simeq
(3L_p\xi_m^2/2)^{1/3}$, are perturbed by the network. For the snake-like regime
we thus identify the entanglement length $L_e^\infty$ for an infinitely long
polymer with $\lambda$. We expect the qualitative features of the entanglement
transition derived above for the rod-like regime to hold also in the snake-like
regime. The isotropic entanglement volume $V_e\simeq\xi_e^3$ in Eq.\ 
(\ref{ansatz}) has now to be replaced by $V_e\simeq\xi_eR^{\perp2}_{L_e}$.  The
implicit dependence of $V_e$ on $L_p$ again reflects the broken scale
invariance in the snake-like regime. We leave the problem of the crossover
between the snake-like and the scale invariant cases for further investigation.

Now we turn to the comparison of our results with available experimental data.
We suppose that the existence of a disentangled phase above the overlap 
concentration $\bar c$, as predicted by Eq.\ (\ref{ansatz}), is likely to 
explain some discrepancies
of the Doi-Edwards theory \cite{doi92,doi75} for the rotational diffusion of
rigid rods in a semidilute solution with experimental data \cite{pec85}. The
Doi-Edwards theory is based on the assumption that for concentrations larger
than the overlap concentration ${\bar c}$ there is a separation of time scales.
One anticipates that each step of the rotational diffusion process is
determined by the constraint that each rod is confined to remain within an
angular range $\xi_T/\!L$ during the time it takes a rod to diffuse a distance
equal to its length. The tube of radius $\xi_T \propto 1/cL$ is assumed to
impose a long-lived topological restriction on the motion of the rods.
However, as we have argued above, long-lived entanglements emerge only at a
much higher concentration $c^\star = 27 {\cal C}\bar c/4$. This would explain
why the onset of entanglement as defined by a marked decrease in the rotational
diffusion coefficient occurs at a concentration $c_{\rm exp}$ fully two orders
of magnitude above the overlap concentration \cite{pec85}.

As an important practical application of the above ideas we already mentioned
actin, a semiflexible macromolecule, which is of major biological interest but
is also an almost ideal model system for physicists \cite{kas95}. It is well
suited to test our ideas, because it is characterized by a large ratio $L_p/a$
($\simeq 10^3$) and thus by a broad semidilute regime, so that the wormlike
chain model applies without modification over several orders of magnitude in
concentration. The average length of the molecules can be adjusted by adding so
called actin binding proteins such as gelsolin or severin.  Existing data on
actin \cite{mul91,jan94,jan88} give only an incomplete picture of the rather
complex situation sketched above but seem to support our results.  For short
(rod-like) filaments a length dependence of the plateau modulus near the
entanglement transition has been observed \cite{jan94}, which is qualitatively
well described by Eq.\ (\ref{G0}) and (\ref{Le}), but is somewhat smeared out
(probably as an effect of sample polydispersity). In the entangled phase Eq.\ 
(\ref{G0}) together with Eq.\ (\ref{Le}) predicts a plateau modulus $G^0\propto
\kappa c^2$ for the rod-like case far from the entanglement transition. Near
the transition the concentration dependence is enhanced according to Eq.\ 
(\ref{Le}). In the snake-like case as defined above we have $L_e\simeq
(L_p\xi_m^2)^{1/3}$ with $\xi_m\propto c^{-1/2}$ and hence $G^0\propto
c^{5/3}\kappa^{1/3}(k_BT)^{2/3}$. In experiments with actin the exponent of the
observed power law for $G^0(c)$ ranges from $1.7$ to $2.3$ in the entangled
regime \cite{mul91,mac95,jan88}. As a critical test of our ideas we suggest a
comparison of the plateau modulus for actin networks with and without
tropomyosin, which is known to cause a considerable stiffening of actin
filaments. The rod-like regime and the snake-like regime should be readily
discernible due to their markedly different dependence of $G^0$ on $\kappa$.
\newline In summary, we have derived the force-extension relation for a
wormlike chain and discussed some of its consequences for the viscoelastic
properties of entangled solutions and gels of semiflexible polymers.
Especially, we analyzed the entanglement transition and predicted various
exponents for the dependence of the plateau modulus $G^0$ on concentration and
bending rigidity. 
\newline It is a pleasure to acknowledge helpful discussions
with Jan Wilhelm, Markus Tempel and Erich Sackmann. We are also thankful to the
authors of Ref.\ \cite{mac95} for making a preprint of their work available to
us. Our work has been supported by the Deutsche Forschungsgemeinschaft (DFG)
under Contract No.\ Fr.\ 850/2 and No.\ SFB 266.

\end{multicols}


\begin{thebibliography}{99}
  
\bibitem{Geilo95} See e.g.: Proceedings of the NATO Advanced Study Institute on
  {\em Physics of biomaterials: fluctuations, self assembly and evolution}
  (Geilo, Norway, 1995).
  
\bibitem{nel95} D. R. Nelson, in {\it Observation, Prediction and Simulation
    of Phase Transitions in Complex Fluids }, edited by M. Baus et al. (Kluwer
  Acad. Publ., Netherlands, 1995).

\bibitem{sac94} E. Sackmann, Macromol. Chem. Phys. {\bf 195}, 7  (1994).
  
\bibitem{fer80} J. D. Ferry, {\em Viscoelastic Properties of Polymeric
    Liquids} (Wiley, New York, 1980).

\bibitem{doi92} M. Doi, S. F. Edwards, {\em The Theory of Polymer Dynamics}
  (Clarendon, Oxford, 1992).

\bibitem{col90} R. H. Colby, M. Rubinstein,  Macromol. {\bf 23}, 2753
(1990). 
 
\bibitem{schweizer} See e.g.: K. G. Schweizer, G. Szamel, Transp. Th.  and
  Theor. Phys. {\bf 24}, 947 (1995).

\bibitem{mul91} O. M\"uller {\em et al.}, Macromol. {\bf 24}, 3111 (1991).
 
\bibitem{jan94} P. A. Janmey {\em et al.}, J. Biol. Chem. {\bf 269}, 32503 
(1994).

\bibitem{zan95} K. S. Zaner, Biophys. J. {\bf 68} 1019 (1995).
  
\bibitem{sai67}See e.g.: N. Saito {\em et al.}, J. Phys. Soc. Jpn.  {\bf 22},
  219 (1967).
 
\bibitem{comple}For strong forces see: C. Bustamente {\em et al.}, Science {\bf
    265}, 1599 (1994); J. F. Marko, E. D. Siggia, Macromol. {\bf 28} 8759
  (1995); T. Odijk, ibid. {\bf 28} 7016 (1995); J. Wilhelm, E. Frey,
  unpublished.
  
\bibitem{mac95} F. C. MacKintosh, J. K\"as, P. A. Janmey, Phys. Rev. Lett.
{\bf 75}, 4425 (1995). The first non-vanishing term in a moment expansion of
  $\tilde f^{-1}$ for a polymer with its end-to-end vector strictly parallel to
  the applied force, namely $\left(\langle R_L^4 \rangle-\langle
    R_L^2\rangle^2\right)/4\langle R_L^2\rangle k_BT$, has exactly the
  asymptotic form derived by these authors.

\bibitem{gib88} L. J. Gibson, M. Ashby, {\em Cellular Solids: Structure \&
    properties} (Pergamon, Oxford, 1988).
  
\bibitem{kre95} At this point an assumption about the network structure and the
  nature of entanglements has to be made. How are macroscopic strains and
  stresses mediated by the network to the individual strands?  The isotropic
  average seems to us the simplest plausible assumption, here. For the case of
  flexible polymers sophisticated network models have been discussed in the
  literature over the last decades; see e.g.\ R.  Everaers, K. Kremer,
  Macromol. {\bf 28}, 7291 (1995).

\bibitem{kav87} T. A.  Kavassalis, J. Noolandi, Phys. Rev. Lett. {\bf 59},
  2674 (1987); Macromol. {\bf 21}, 2869 (1988); ibid. {\bf 22}, 2709
  (1989).
 
\bibitem{odi83} T. Odijk, Macromol. {\bf 16}, 1340 (1983).

\bibitem{footn} In Ref. \cite{kav87} it was argued that ${\cal C}$ is a
  universal quantity. From an analysis of experimental data the authors
  calculate a mean value ${\cal C} = 9.1 \pm 0.8$.  In the case of random coils
  ${\cal C}-1$ should be equal to the ``coordination number'' $K$ because of
  their isotropic conformations. But in the rigid rod limit $K$ has to be
  larger than ${\cal C}-1$, because a polymer crossing the test volume will in
  general not go through the center of the sphere. Elementary geometrical
  considerations suggest that in this case the coordination number is
  $K=16({\cal C}-1)/3\pi$. Hence for a given value of ${\cal C}$ the number of
  nearest neighbors involved in an entanglement is larger for semiflexible
  polymers as compared to random coils.  This is in accord with ones intuition
  that flexible coils, resembling interpenetrating fuzzy balls, are more
  efficient in mutual entanglement than semiflexible polymers, which are
  rod-like on short scales.

\bibitem{doi75} M. Doi, J. Physique {\bf 36}, 607 (1975); M. Doi, S.F.
 Edwards, J. Chem. Soc.  Far. Trans. II {\bf 74}, 560, 918 (1978).

\bibitem{pec85} R. Pecora, J. Pol. Sci.: Pol. Symp. {\bf 73}, 83 (1985); and
references cited therein.

\bibitem{kas95} J. K\"as {\em et al.}, Biophys. J., {\bf 70}, 1 (1996).

\bibitem{jan88} P. A. Janmey {\em et al.}, Biochemistry {\bf 27}, 8218 (1988).

\end{thebibliography}
\end{document}